# Cavity Ring-down UV spectroscopy of the $C^2\Sigma^+$-$X^2\Pi$ electronic transition of CH

Chris Medcraft[1*], Harold Linnartz[2], Wim Ubachs[1]

[1]Department of Physics and Astronomy, LaserLaB, Vrije Universiteit Amsterdam,
De Boelelaan 1081, NL-1081 HV Amsterdam, The Netherlands

[2]Sackler Laboratory for Astrophysics, Leiden Observatory, Leiden University,
PO Box 9513, NL-2300 RA Leiden, the Netherlands

**Abstract**

Rotationally resolved spectra of the $C^2\Sigma^+$-$X^2\Pi$ electronic system of the CH radical were measured using cavity ring-down spectroscopy in supersonically expanding, planar hydrocarbon plasma. The experimental conditions allowed the study of highly excited rotational levels starting from vibrationally excited states. Here we present some 200+ new or more accurately recorded transitions in the 0-0, 1-1 and 2-2 vibronic bands in the ultraviolet between 30900-32400 cm$^{-1}$ (324-309 nm). The resulting data, compared to earlier measurements, allows for the determination of more precise molecular constants for each vibrational state and therefore more precise equilibrium values. From this an equilibrium bond length of 1.115798(17) Å for the $C^2\Sigma^+$ state is determined. A comprehensive list with observed transitions for each band has been compiled from all available experimental studies and constraints are placed on the predissociation lifetimes.

**1. Introduction**

The carbon hydride (CH) radical has a long history in molecular spectroscopy. First detected in the laboratory in 1919 [1], it was subsequently among the first positively identified extra-terrestrial molecules [2]. CH can be found in many physical and astrophysical environments; it is for example the cause of blue colour of hydrocarbon flames, where it acts as an important reaction intermediate. In space, the CH radical is found in cometary tails [3], stellar atmospheres [4], protostellar accretion disks [5], diffuse [6] and dense [7] interstellar clouds, as well in extra-galactic sources [8]. These sources span environments exhibiting a large range of thermodynamic properties. This has resulted in many transitions of CH being detected first in astronomical sources before their laboratory detection was established. The Λ-doublet transitions in the ground state were first detected using radio astronomy in 1973 [7]. Rotational constants for the v=4 and v=5 levels of the $A^2\Delta$ state were derived for the first time using spectra from carbon enhanced metal poor stars [4]. These spectra also showed transitions of higher rotational quanta than were recorded in the laboratory for the $B^2\Sigma^+$- $X^2\Pi$ and $C^2\Sigma^+$- $X^2\Pi$ band. Notably the $A^2\Delta$-$X^2\Pi$ transition contributes to the Fraunhofer G-band that is used to study stars in a wide range of stellar types. Currently the most precise molecular parameters for the ground $X^2\Pi$ electronic state use transitions observed in solar absorption spectra recorded using the ACE-FTS instrument [9]. The optical absorptions of CH, and other diatomics like $C_2$ and CN can be used to classify carbon rich stars [10] providing crucial information on the chemistry and radiation transfer in stellar atmospheres [11]. The CH spectrum is also an important tool in measuring $^{12}C/^{13}C$ ratios in space.

[*]Corresponding author.
Present address: School of Chemistry, UNSW Sydney, Sydney, New South Wales 2052, Australia
email: c.medcraft@unsw.edu.au

Accurate laboratory spectroscopic information over the widest range of rotational and vibrational quanta is needed over all wavelength ranges for further constraining astronomical models and aiding observations [12]. The Λ-doubling in the ground state has been studied at ever growing accuracy by Bogey et al [13], by Brazier and Brown, [14] by McCarthy et al. [15] and ultimately by Truppe et al. [16,17] who measured transitions with an accuracy of 3 Hz. This accuracy, when combined with astronomical observations, can put limits on a possible variation of fundamental constants for which the ground $X^2\Pi$ state is particularly sensitive [18]. The work by Gerö [19] reported rotationally resolved spectra of the optical and UV bands from the ground state to the $A^2\Delta$, $B^2\Sigma^-$ and $C^2\Sigma^+$ states. The A-X bands were studied further at higher precision by Brazier and Brown [14] and by Ubachs et al. [20]. The 0-0 vibronic band of the C-X transition was first studied by Heimer [21], then by Herzberg and Johns [22] who included measurements on the 1-1 and 2-2 bands. More precise measurements on the $C^2\Sigma^+$-$X^2\Pi$ 0-0 band were made by Ubachs et al. [23] and an extended line list of this band was produced by Bembenek et al. [24]. More precise observations of the 1-1 band were reported by Li and co-workers [25].

These existing data and line lists (i.e. overview of all observed and/or calculated transition frequencies) for the lower electronic states of CH ($X^2\Pi$, $A^2\Delta$, $B^2\Sigma^-$ and $C^2\Sigma^+$) were critically analysed and summarised by Masseron and co-workers [4] who combined data from many laboratory and astronomical sources to produce sets of self-consistent line lists and molecular constants for astronomical observations. We use their work as the basis for our assignments and fitting of a large number of new transitions in the UV $C^2\Sigma^+$-$X^2\Pi$ electronic band system. The vibrational levels of the $X^2\Pi$ ground state have been studied up to the v=5 level; here we fix the rotational constants for the $X^2\Pi$ state to those presented in Masseron et al. [4] who primarily used data from Colin and Bernath [9]. In addition to the experimental studies there are several high-level *ab initio* studies on the valence and Rydberg states [26–28]. Theoretical work on the potential energy curves of CH have predicted that the $C^2\Sigma^+$ state interacts strongly with the yet unobserved $2^2\Sigma^+$ state [26,27].

## 2. Experimental

UV spectra of CH were recorded using pulsed cavity ring-down (CRD) spectroscopy [29]. The frequency-doubled output of a tuneable dye laser (SIRAH Cobra), running on DCM, was used to produce UV pulses (≈0.08 cm$^{-1}$ bandwidth) which were coupled to an optical cavity with high reflectivity mirrors (R≈99.9%). Each pulse of light was confined within the cavity and the exponential decay of the ring-down was measured with a photomultiplier tube and oscilloscope. For a cavity length of 52 cm typical ring-down times amounted to 1.7 μs. This relatively short ring-down time was determined by the presently available quality of high reflectivity mirrors in the UV range. The red output of the dye laser was used to simultaneously record etalon and iodine ($I_2$) transmission spectra for calibration purposes to achieve an absolute accuracy of each data point of *ca.* 0.02 cm$^{-1}$.

The molecular radicals were produced in a plasma slit jet source that has been described in detail elsewhere [30,31]. Briefly, a high-pressure (≈6 bar) mixture of acetylene (≈0.3%) diluted in argon is pulsed into a vacuum chamber by a solenoid valve (General Valve, Series 9) mounted in front of the slit with *ca.*1 ms opening duration. A 1000 m$^3$/hr roots blower system keeps pressures below 0.1 mbar during jet operation. A high negative voltage (≈ -1000V) is discharged between electrodes isolated by ceramic spacers with slit openings of 3 cm × 500 μm. The discharge occurs over 300− 500 μs to create a planar plasma which dissociates the acetylene and facilitates molecule formation through collisions in the expanding plasma. This source has been shown to be capable of producing long chain hydrocarbon radicals such as $C_6H$ [32] or $C_9H_3$ [33] and supersonically cool them to rotational temperatures as low as 15-25 K. Here we used this source to study the much lighter CH fragment, for which a less effective cooling is expected. Labview routines were used to guarantee that cavity ring-down and plasma pulse coincide and for data acquisition. The plasma jet was used at 10 Hz and typically 15 ring-down events were averaged to generate one data point.

## 3. Results

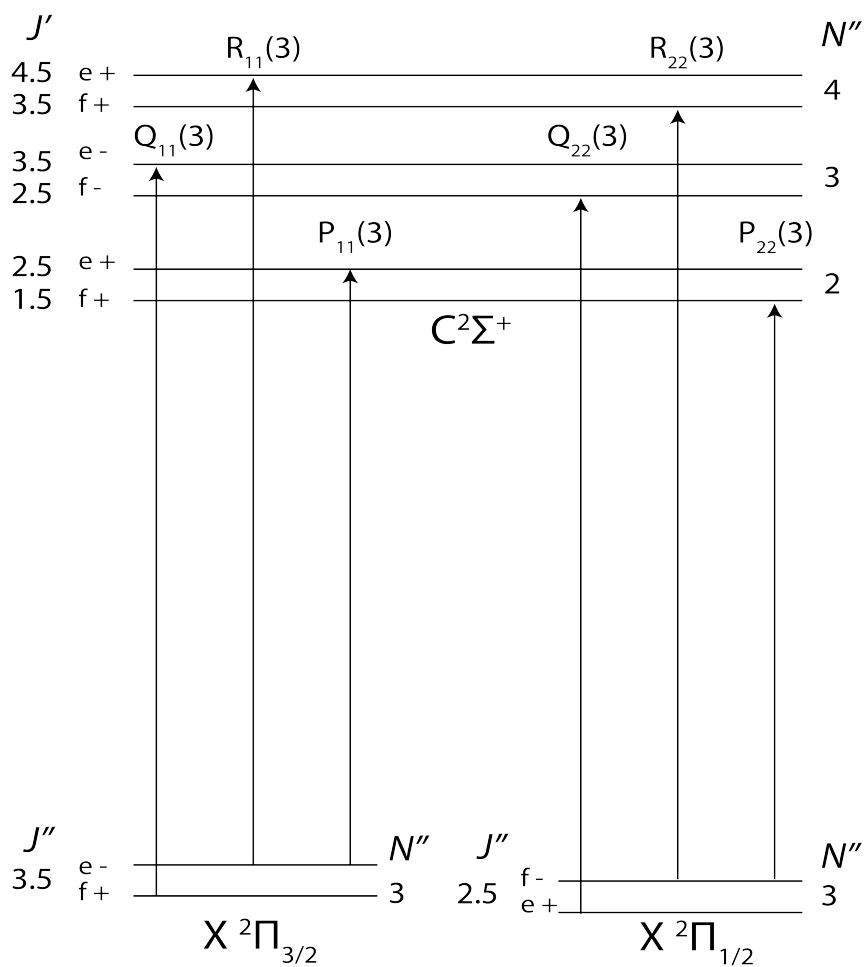

Figure 1: Schematic illustrating the six rovibronic bands of the $C^2\Sigma^+$-$X^2\Pi$ band of CH for the chosen energy levels, e/f refer to electronic symmetry, +/- refer to total parity.

The $C^2\Sigma^+$-$X^2\Pi$ vibronic bands of CH are composed of six main rotational branches $P_{11ee}$; $P_{22ff}$; $Q_{11ef}$; $Q_{22fe}$; $R_{11ee}$ and $R_{22ff}$ as shown in Fig 1. A further six satellite bands $P_{12ee}$; $Q_{12ef}$, $R_{12ee}$; $P_{21ff}$; $Q_{21fe}$, and $R_{21ff}$ are also allowed although much weaker. The three observed vibronic bands are 0-0 ($T_{00}$=31792 cm$^{-1}$), 1-1 ($T_{11}$=34403cm$^{-1}$) and 2-2 ($T_{22}$=36773cm$^{-1}$). The Q-branch region for each band is shown in Fig. 2, where the N quantum numbers of assigned transitions are labelled above the spectra. The nearly equal observed intensities of the 0-0 and 1-1 transitions imply a vibrational temperature of at least 5000 K. Rotational temperatures also appear to be very high with observed transitions starting in the 0-0 band from N-values as high as 30.

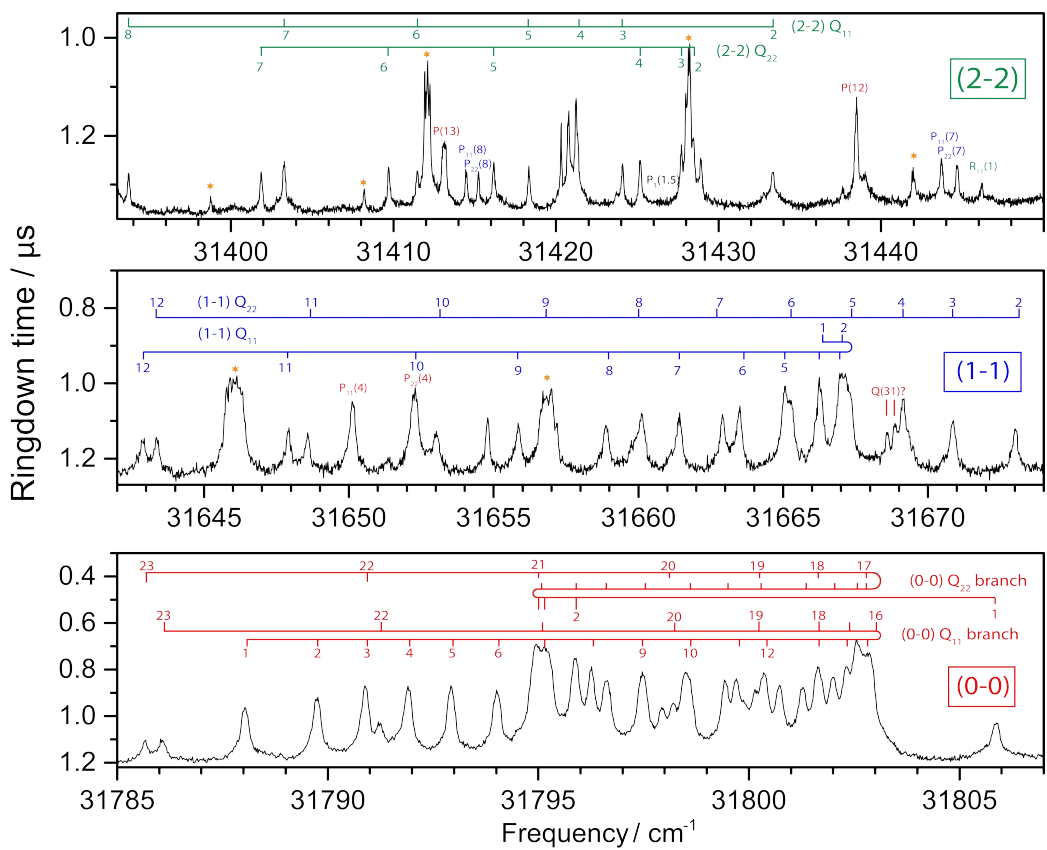

Figure 2: Q-branch regions of the 0-0 band (bottom panel, assignments in red), 1-1 band (middle, blue) and 2-2 band (top, green). Lines from the 0-0 band are labelled in red in the middle and upper panels. Lines from the 1-1 are labelled in blue in the upper panel. The features indicated with stars are of unidentified origin.

### 3.1 Temperature distributions

The observed rotational intensity pattern shown in Fig. 2 cannot be reproduced with a standard Boltzmann distribution. For example, the intensities of $Q_{11}$ transitions from $N$=1 to $N$=6 (31788-31794 cm$^{-1}$) in the 0-0 band shown in the lower panel of Fig. 2 can be approximated with $T_{rot}$≈500 K. However, at this temperature transitions of $N$=9 have a predicted intensity more than 10 times less than $N$=3 and for 500 K the observed band head at 31803 cm$^{-1}$ cannot be reproduced at all. To reproduce the observed intensities of the transitions involving higher rotational states a much higher rotational temperature (*ca.* 5000 K) is required. At this temperature the transitions involving lower rotational states are predicted to have much lower intensities than observed. The non-Boltzmann conditions in the slit jet expansion can be explained by the large energy gap between rotational levels in CH and by the number of inelastic collisions with argon in the expansion not being sufficiently large to efficiently cool the plasma excited species. Cooper et al. [34] showed that the rotational energy relaxation of CH (in the A$^2$Δ state) by collisions with argon are driven primarily via transitions of Δ$N$=+/-1 and that rates are highest at $N$=4 and drop by a factor of ≈100 at $N$=25 [34]. Thus, we presume that the CH radical is formed in a very highly rotationally and vibrationally excited state inside the plasma. While some lower rotational states are cooled, the energy gap between the higher states is too large and the relaxation rate is too slow and thus significant population remains at high rotational states. Similarly, the vibrational cooling is inefficient due to the large energy difference between vibrational levels (≈2800 cm$^{-1}$) of the X$^2$Π electronic state resulting in substantial populations in the v=1 and v=2 vibrational states.

Changes made to experimental conditions of the plasma and expansion generally support this supposition. Reducing backing pressure of the acetylene argon mixture from 6 bar to 1 bar did not change the relative intensity of rotational lines significantly (Fig 2. lower panel). Likewise increasing the distance between the discharge source and the optical axis appears only to have a minor influence. When the valve is moved 4.7 cm away from the optical axis, thus allowing for continuing collisional cooling, a slight increase is observed in the lower rotational transitions and a small decrease is seen for transitions leaving from higher states (Fig 2. upper panel). For an argon-dominated supersonic expansion this corresponds to measuring the molecules after ≈84 µs of free expansion. One would normally expect these variables to change the observed rotational spectrum, however the *N*=23 transitions are still observable in both cases. This suggests the molecules are either 'trapped' in the rotational states that these transitions originate from or that they are replenished via cooling from higher states. This observation is also consistent with conclusions derived from a plasma expansion study of the $A^2\Sigma^+$-$X^2\Pi$ electronic transition of the heavier CF [35]; also here higher vibrational levels were found to be substantially populated, but rotational temperatures were low as CF can cool much easier, rotationally, given its much smaller rotational constant.

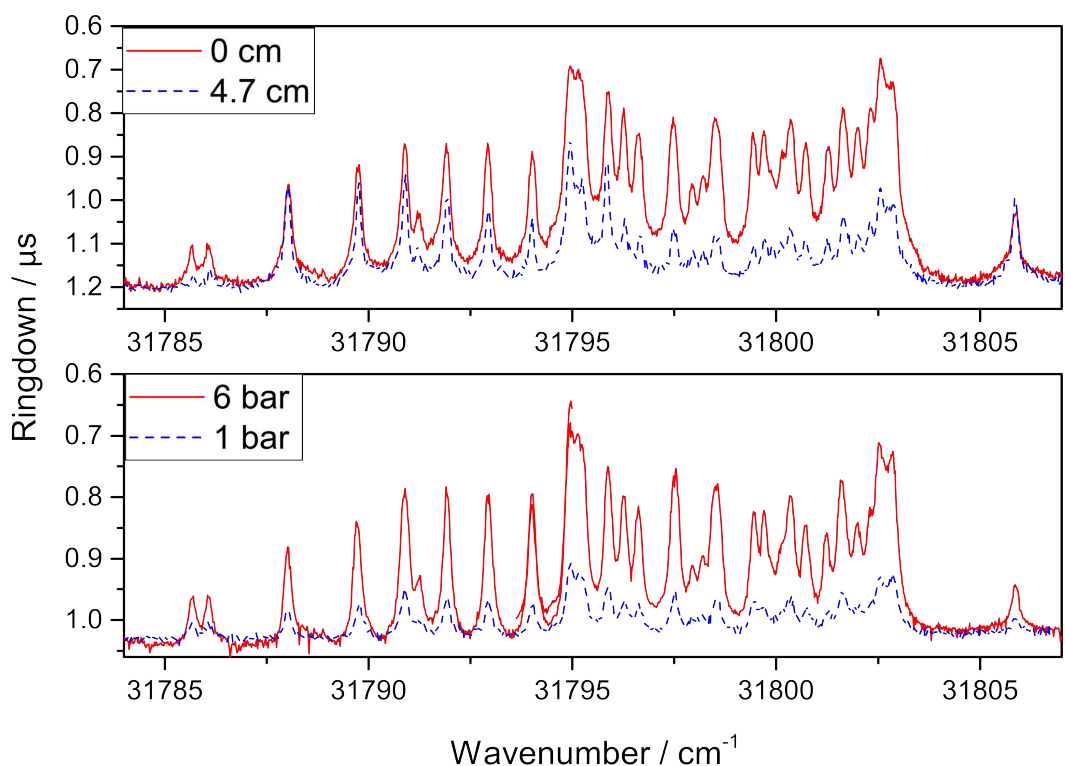

Figure 3: The Q-branch region of the 0-0 band of the C-X transition at different experimental conditions. Top panel compares a CRDS measurement close to the nozzle orifice (red) and 4.7cm down (blue). The bottom panel compares a high backing pressure of $C_2H_2$/Ar of 6 bar (red) with a lower pressure of 1 bar (blue).

### 3.2 Spectral analysis

The population of quantum states in all three vibrational levels v=0,1,2, and up to large rotational angular momenta in the slit jet expansion was employed here provided information on a wide range of levels in the $C^2\Sigma^+$ excited state. For the spectral fitting procedure, first each vibronic band was simulated in PGOPHER [36] using rotational constants for each vibrational state of the $X^2\Pi$ and $C^2\Sigma^+$ levels as presented in Masseron et al. [4]. We kept the constants for the $X^2\Pi$ ground state fixed and used them to compute lower state combination differences to confirm our, and previous, line assignments. Using this simulation as a starting point, line positions from former studies were entered into the line list in PGOPHER with appropriate uncertainties (see below). For each vibronic band we extended the literature line list by adding our newly measured transitions or replaced line positions wherever our data has a greater estimated precision than that of previously measured transitions. In total 348 transitions from six studies were included in the fits, comprising as many as 216 new or improved line positions as derived from the work presented here. These line lists were then used to perform a weighted least-squares fit in PGOPHER of the upper state ($C^2\Sigma^+$), while keeping the lower state constants fixed to the numbers available from Masseron et al. [4]. This yields the excited state constants listed in Table 1.

### 3.3 0-0 band

For the 0-0 band there are three relevant sets of data from previous measurements. The most extensive line list is from Heimer [21] which has lines up to $N$=28 that were measured with an estimated precision of approximately 0.2 cm$^{-1}$ [4,25], about 4 times less precise than in the present work. These lines are included only in our fit when no data from other measurements is available. All lines from the high-resolution study by Ubachs et al. [23] are included as they have the most precise line positions (≈0.004 cm$^{-1}$), although only lower rotational energy levels were measured. A more recent study by Bembenek et al. [24], performed in emission, reported line positions for 63 transitions. However due to a large background emission only 22 lines were included in their fit [24]. In total we measured 73 lines up to $N$=30 with an estimated precision of 0.05 cm$^{-1}$. The fit presented in Table 1 also includes 28 lines from Heimer [21], 35 from Ubachs et al. [23] and 22 from Bembenek et al. [24], all weighted according to their estimated uncertainties. The full line list is presented in Table 3, added as an Appendix. In order to fit all lines to within experimental uncertainties, inclusion of the $M$ centrifugal distortion constant was required. A recent study on the same band of the CD radical (up to $N$=35) was able to fit the $M$ centrifugal distortion constant to 8.22×10$^{-16}$ cm$^{-1}$ [37], i.e. two orders of magnitude lower than the value fitted here for CH. This difference may be due to a perturbation of the rotational energy levels present in CH but not CD.

A comparison of residuals (observed minus calculated line positions) using constants from Masseron et al. [4] (open triangles) and this work (solid circles) is shown in Fig. 4. The newly observed transitions at higher rotational levels show very high deviations (up to 5 cm$^{-1}$) from the calculated positions of Masseron et al. [4]. While this deviation can be fit by adding the extra centrifugal distortion constant, the large magnitude of the constant and the large deviations without it hint at a possible perturbation to this state. Theoretical studies on CH have indicated that there exists an anti-crossing between the potential energy curves of the $C^2\Sigma^+$ and the (2)$\Sigma^+$ state [28] (also labelled as the $D^2\Sigma^+$ state [27]) which could push the rotational levels down. No anti-crossing could be observed in the rotational levels involved in the detected transitions here and perturbation fits that included an interaction with the (2)$\Sigma^+$/$D^2\Sigma^+$ state did not converge. Nevertheless, it is still suspected that the higher rotational levels of the $C^2\Sigma^+$ state may be perturbed and therefore the fit results presented in Table 1 should be considered as effective parameters; extrapolation beyond the observed data using these constants should be performed with care.

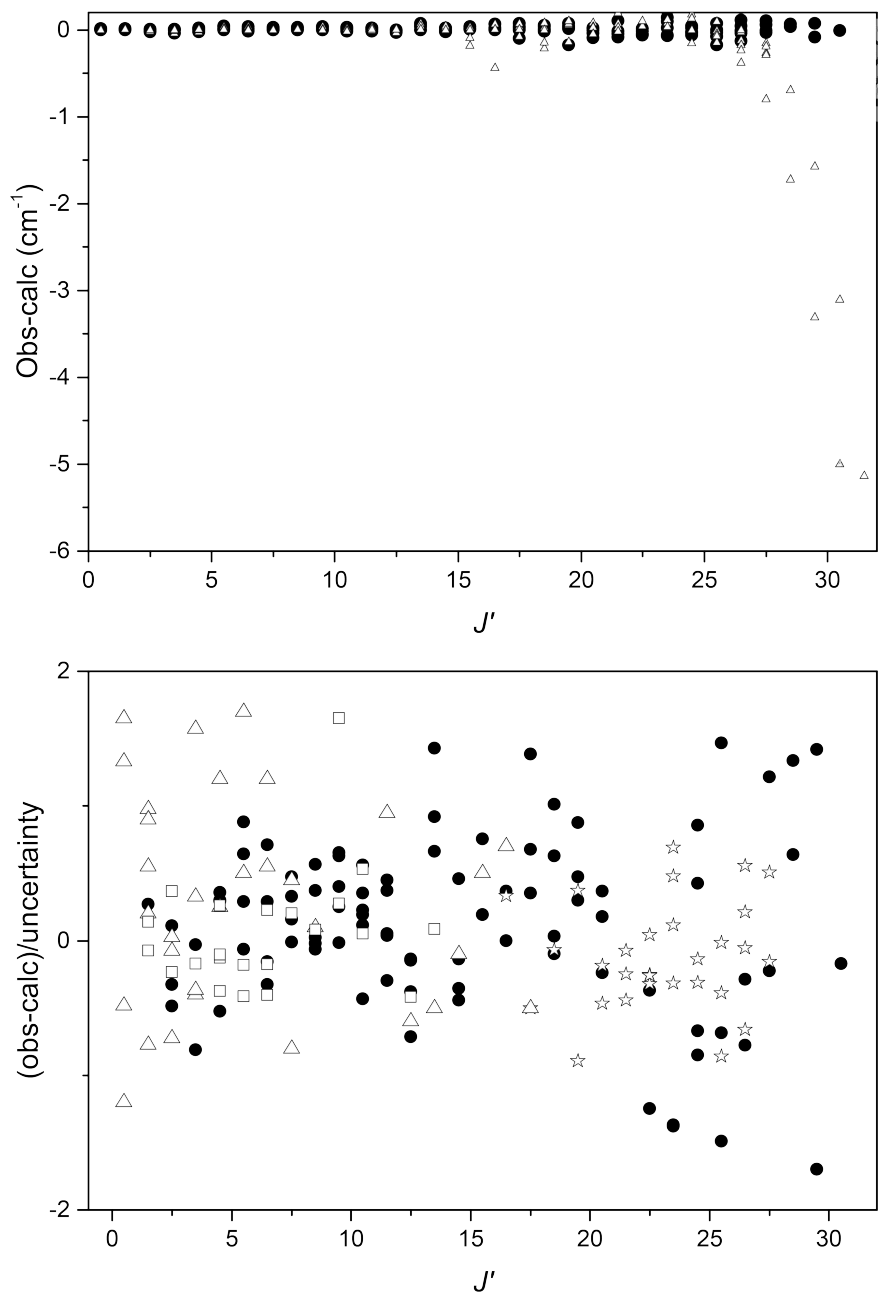

Figure 4: Residual plots for the 0-0 vibronic band. Upper: comparison of residuals of all data presented here when using molecular constants determined here (solid circles) and the constants presented in Masseron et al. [4] (open triangles). Lower: weighted residuals from the fit presented for all lines in this work (solid circles), from Heimer [21] (open stars), Ubachs et al. [23] (open triangles) and from Bembenek et al. [24] (open squares).

### 3.4 1-1 band

For the 1-1 band there is only one previous study with relevant data for inclusion in the present fit [25]. This double resonance experiment used the A-X band as a probe for the $C^2\Sigma^+$ v=1 levels. The estimated precision of the recorded transitions is approximately 0.2 cm$^{-1}$, but as the double-resonance detection scheme employed is state selective, overlapping transitions could be resolved. The extensive line list presented in previous work [4] could not be extended here, but a total of 83 transitions were measured with a higher precision (0.05 cm$^{-1}$) and were combined with 46 transitions from Li et al. [25]. The full line list can be found in Table 4 (see Appendix). While the original fit presented by Li et al. [25] included the $L$ centrifugal rotational constant the fit presented by Masseron et al. [4] omits this constant which results in the sign of the $H$ constant changing. The higher precision of the lines in the present work required fitting the $L$ distortion constant to fit all observations to within experimental uncertainty. Without it the rms error of the fit is twice as large as that presented in Table 1. The fitted value of the $D$ distortion constant in this work ($1.7097(30) \times 10^{-3}$ cm$^{-1}$) is larger than the previous result ($1.6336(21) \times 10^{-3}$ cm$^{-1}$) [4].

### 3.5 2-2 band

The only previous laboratory measurements of the 2-2 band are from Herzberg and Johns [22] who included lines up to $N=8$. In the present study we were able to measure 62 lines up to $N=16$ while only two lines from the previous study were included in our fit. The line list for the 2-2 band is presented in Table 5 (see Appendix). These new measurements increased the precision of the v=2 rotational constants by almost an order of magnitude compared to previously reported constants. The improved parameters are important in the determination of equilibrium parameters (section 3.5). Given the vast improvements in technology available to molecular spectroscopists, it is somewhat remarkable that it has taken some 50 years to improve upon the earlier laboratory measurements of this specific band. This is perhaps due to the difficulty in preparing large densities of molecules in such highly excited vibrational and rotational states, thus showing the utility of the source used in this work for spectroscopy on highly excited levels of small molecules. Figure 5 shows a small section of the spectrum with transitions from each of the vibronic bands displayed. This clearly shows once again the near equal intensity of each band. The intensity of the 2-2 band and a reasonably high Franck-Condon factor of 0.95 [38] prompted the search for the unobserved 3-3 band and although there are many unidentified features in the spectra none of these could be positively assigned to the 3-3 band.

Over all three vibronic bands 216 transitions were measured either for the first time or with increased precision. These were combined with previous measurements to produce improved molecular constants for the v=0, 1 and 2 vibrational states of $C^2\Sigma^+$ which are shown in Table 1 and compared to the constants from Masseron et al. [4]. It is clear that the constants compare very well, but that the larger data set presented here allows for further improved results.

Table 1: Rotational constants for the v=0,1,2 vibrational levels of the $C^2\Sigma^+$ state of CH (in cm$^{-1}$).

| Parameter | v=0 | v=1 | v=2 |
|---|---|---|---|
| $T_v$ | 31791.64843(64) | 34403.2054(83) | 36772.8964(94) |
| | *31791.64790(160)* | *34403.310(21)* | *36772.842(54)* |
| $B_v$ | 14.255562(36) | 13.51436(32) | 12.60583(44) |
| | *14.255724(90)* | *13.50577(43)* | *12.6046(29)* |
| $\gamma_v$ (x10$^2$) | 4.101(15) | 3.545(66) | 3.55(10) |
| | *4.126(58)* | *3.320(120)* | *3.80(10)* |
| $D_v$ (x10$^3$) | 1.59010(47) | 1.7097(30) | 1.9354(45) |
| | *1.59375(97)* | *1.6336(21)* | *1.918(37)* |
| $H_v$ (x10$^8$) | 5.27(20) | 8.015(998) | -68.8(1.2) |
| | *7.45(29)* | *−14.7(27)* | *−73.(10)* |
| $L_v$ (x10$^{11}$) | 1.40(32) | -2.17(11) | |
| | *−3.75(26)* | --- | |
| $M_v$ (x10$^{14}$) | -3.88(17) | | |
| | --- | | |
| $\gamma_D$ (x10$^5$) | -1.198(89) | | |
|                    | *−1.66(62)* |           |           |
| No. of transitions | 158       | 129       | 62        |
| σ                  | 0.81      | 0.63      | 0.64      |
| $N_{max}$          | 30        | 23        | 16.5      |

Roman type numbers are constants determined in this work including new and more precise line positions, italic numbers are from Masseron et al. [4], all values in cm$^{-1}$, numbers in parenthesis represent one standard deviation.

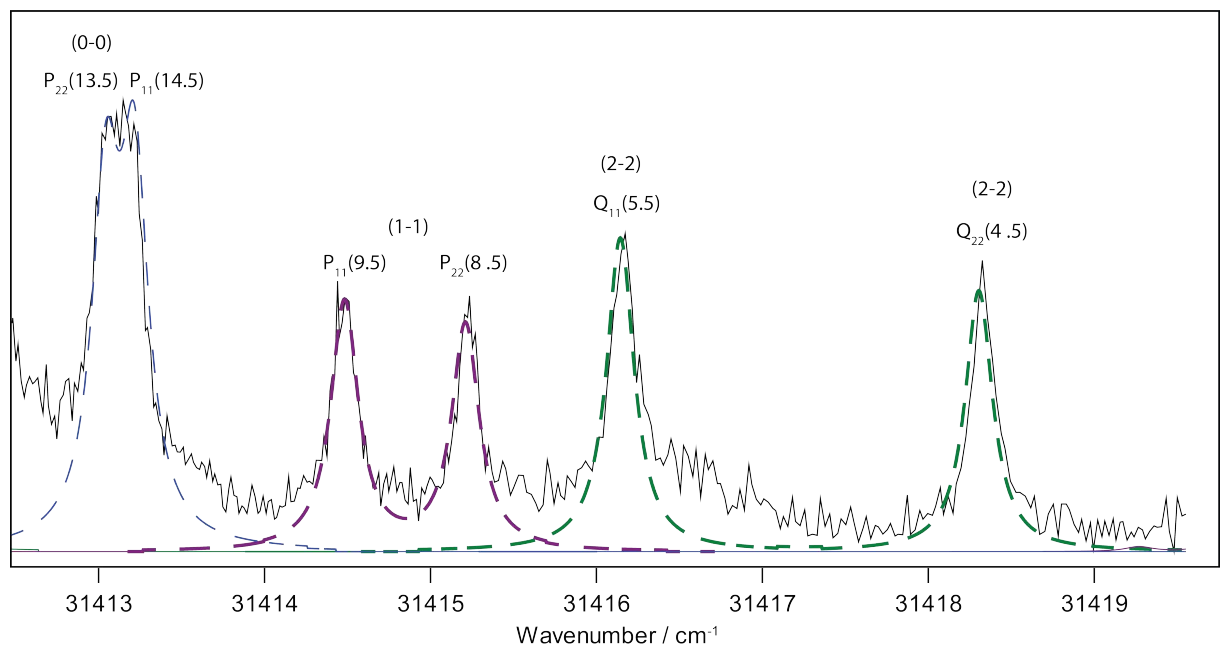

Figure 5: A zoom of the Q-branch region of the 2-2 band of the $C^2\Sigma^+$-$X^2\Pi$ transition of CH showing the line profiles of transitions from all three vibronic bands with near equal intensity and line width in a single recording. Simulated line widths (dashed lines) are Lorentzians with 0.2 cm$^{-1}$ FWHM as simulated in PGOPHER.

### 3.5 Equilibrium constants

From the rotational constants of the $C^2\Sigma^+$ state (Table 1) and the previously determined constants for the ground electronic state, equilibrium rotational constants can be calculated assuming:

$$B_v = B_e - \alpha_e \left(v + \frac{1}{2}\right) + \gamma_e \left(v + \frac{1}{2}\right)^2$$

where $B_e$ and $B_v$ are the rotational constants for the equilibrium and the v=0,1,2 vibrational states, respectively. For a diatomic molecule the equilibrium rotational constant directly gives the equilibrium bond length and the fundamental frequency ($\omega_e$), anharmonic constant ($\omega_e x_e$) and vertical transition energy ($T_e$) can be calculated. These are presented in Table 2 with a comparison to those found in Masseron et. al. [4] and *ab initio* values from Kalemos et al. [27]. The theoretical values are in good agreement with those determined here. The equilibrium rotational constant ($B_e$) and the vibration-rotational interaction constants ($\alpha_e$ and $\gamma_e$) are in reasonable agreement with Masseron et al. [4]. Note that the equilibrium constants reported in Masseron et al. [4] were presented without uncertainties. The higher uncertainties in the rotational constants (Table 1) should propagate to proportionally higher uncertainties for the equilibrium constants (Table 2).

Table 2: Equilibrium constants for the $C^2\Sigma^+$ state of CH.

| Constant | This work[a] | Masseron[b] | Kalemos[c] |
|---|---|---|---|
| $B_e$ (cm$^{-1}$) | 14.5634(4) | 14.56066 | 14.5477[d] |
| $D_e$ ($10^{-4}$cm$^{-1}$) | 15.701(20) | | 14.91 |
| $\alpha_e$ (cm$^{-1}$) | 0.5741(11) | 0.56653 | 0.429 |
| $\gamma_e$ (cm$^{-1}$) | -0.0836(4) | –0.08659 | |
| $\omega_e$ (cm$^{-1}$) | 2853.423(27) | 2853.1724 | 2837.3 |
| $\omega_e x_e$ (cm$^{-1}$) | 120.9330(95) | 120.8564 | 87.76 |
| $r_e$ (Å) | 1.115798(17) | 1.12777 | 1.1164 |
| $T_e$ (cm$^{-1}$) | 31809.536(14) | 31809.6428 | 32125.33[e] |

a: one standard deviation in parentheses;
b: from Masseron et al. [4];
c: *ab initio* values from Kalemos et al. [27] (CASSCF+1+2);
d: calculated from value of $r_e$;
e: converted from eV.

### 3.6 Predissociation

In early spectroscopic studies Norrish et al. [39] had observed that the C-X band in CH is relatively weak in emission, while it is the strongest in absorption. The work of Herzberg and Johns [22] on CD and CH revealed that the emission from CD is up to 5 times stronger than that of CH. Hesser and Lutz [40 directly measured a difference in the non-radiative lifetimes between these two isotopic species. Together these observations suggested that the $C^2\Sigma^+$ state is predissociated via coupling with another electronic state, causing a mass-dependent and rotational-state dependent predissociation rate. Theoretical calculations of predissociation rates by Van Dishoeck [26] showed that predissociation occurs primarily via spin-orbit coupling to the $B^2\Sigma^-$ state and produced a radiative lifetime of 85 ns. In combination with measurements of the experimental lifetime of the $C^2\Sigma^+$, v=0 level, yielding 3.7 ± 1.0 ns for upper spin-rotational components ($F_1$) and 8.0 ± 1.5 ns for the lower spin-rotational ($F_2$) components [23], this implies a dissociation probability of 90% for the v=0 level. The measurements of Li et al. on the v=1 level constrained the line broadening due to predissociation for $N$=23 to be less than 0.1 cm$^{-1}$, thus the lifetime must be longer than 50 ps.

In the present study we observed linewidths of 0.20 ± 0.04 cm$^{-1}$ for well isolated lines with no discernible change over higher rotational or vibrational states. This is illustrated in Fig. 5, showing recordings of transitions probing all three vibrationally excited states $C^2\Sigma^+$, v=0, 1 and 2 in a single measurement. The solid line in Fig. 5 is the experimental spectrum, whereas the dashed lines represent simulated line positions based on the constants shown in Table 1 with a 0.2 cm$^{-1}$ FWHM Lorentzian line profile. This lineshape is produced from contributions by the laser bandwidth (0.08 cm$^{-1}$ in the UV), Doppler broadening and the natural lifetime associated with predissociation. In view of the fact that excited states of $C^2\Sigma^+$, v=0, known to exhibit a lifetime of about 5 ns, are also broadened to 0.2 cm$^{-1}$, we conclude that the 0.2 cm$^{-1}$ width is entirely instrumental, and a result from laser bandwidth and Doppler width. The Doppler broadening in the present slit-jet discharge expansion experiment was somewhat less than in the experiment of Li et al [25], where constraints were given for lifetimes of $C^2\Sigma^+$, v=1 levels.

Based on the measurements we can extend the information on lifetime broadening of the $C^2\Sigma^+$ state. In $C^2\Sigma^+$, v=0 rotational levels $N$=28-30 were probed for the first time. In view of the low signal-to-noise ratio lifetime broadening can only be constrained to less than 0.1 cm$^{-1}$, indicating that their lifetimes exceed 50 ps. For the $C^2\square^+$, v=1 level we can limit the broadening to <0.05 cm$^{-1}$ for transitions to $N$≤11, hence lifetimes exceeding 100 ps. For the C$^2\Sigma^+$, v=2 level we find quantitative information on predissociation for the first time: lifetimes for $N$≤8 are estimated to exceed 100 ps, and for rotational levels up to $N$=16 to exceed 50 ps.

**Conclusions**

A pulsed high voltage discharge molecular source was used to create a planar, supersonically expanding plasma that proved to be a good source of rotationally and vibrationally highly excited CH radicals. The high rotational and vibrational temperatures are explained by the large energy gaps between states and the inefficient cooling of the argon expansion after plasma formation. This was spectroscopically useful as it allowed the high vibrational and rotational energy levels to be probed by cavity ring-down spectroscopy. This approach may be useful for other small and light molecules, characterized by large vibrational and rotational energy spacings. The recorded spectra were used to extend the experimental line lists of the v=0, 1 and 2 vibrational levels of the C$^2\Sigma^+$ electronic state and derive rotational constants reproducing the fully extended set with transitions. The more precise constants for the v=2 state resulting from the work presented here, assisted in further constraining the equilibrium molecular parameters and predissociation lifetimes. This work along with the recent work on CD [37], provides an excellent basis for molecular modelling and calculations. Although the C-X transition has a similar inherent strength as the A-X transition ($T_{00}$=23173 cm$^{-1}$), the latter has been of more utility for astronomers due to the accessibility and higher radiation flux in the visible spectral region [11]. The work on the C-X system may become beneficial when more sensitive UV telescopes, such as the World Space Observatory-Ultraviolet (WSO-UV) [41] come online or for measurements of sources that are brighter in the UV or where CH is more thermally excited. The bands reported here, also coincide with a wavelength domain in which diffuse interstellar band (DIBs) spectra along many different lines of sight have become available [42] and the search for excited CH transitions may be helpful in further constraining the processes involved in the formation of DIB carriers. In the meantime, the parameters may prove useful as input in radiation transfer models of stellar atmospheres [11,12].

**Supplementary Material**

Data for this work are presented in the form of PGOPHER files.

**Acknowledgement**

The authors thank Carla Kreis and Xavier Bacalla for assistance during part of the measurements. This work has been supported through a grant within the framework of the Dutch Astrochemistry Network funded by NWO (Netherlands Organisation for Scientific Research).

Table 3: Measured transition frequencies of the 0-0 vibronic band of the $C^2\Sigma^+$-$X^2\Pi$ transition of CH (in cm$^{-1}$).

| J' | P$_{11ee}$ | obs-calc | P$_{22ff}$ | obs-calc | Q$_{11ef}$ | obs-calc | Q$_{22fe}$ | obs-calc | R$_{11ee}$ | obs-calc | R$_{22ff}$ | obs-calc |
|---|---|---|---|---|---|---|---|---|---|---|---|---|
| 0.5 | 31759.4753(40)[a] | 0.007 | 31738.6455(20)[a] | -0.0069 | | | 31805.8329(50)[a] | -0.0032 | | | | |
| 1.5 | 31732.5867(40)[a] | 0.0012 | 31709.1003(30)[a] | 0.0001 | 31788.0198(20)[a] | 0.0027 | 31795.8444(20)[a] | -0.0038 | | | 31862.679(20)[b] | 0.0028 |
| 2.5 | 31705.117(20)[b] | -0.0047 | 31680.4677(30)[a] | -0.0008 | 31789.7384(20)[a] | -0.0008 | 31794.931(20)[a] | -0.0034 | 31844.9900(60)[a] | 0.0047 | 31880.952(20)[b] | 0.0073 |
| 3.5 | 31677.5709(50)[a] | -0.0072 | 31652.307(20)[b] | -0.0034 | 31790.8779(40)[a] | -0.0018 | 31794.8994(20)[a] | -0.0019 | 31874.9650(30)[a] | 0.0056 | 31908.0602(20)[a] | 0.0011 |
| 4.5 | 31650.11(20)[b] | -0.0075 | 31624.509(20)[b] | -0.0021 | 31791.9042(20)[a] | 0.0031 | 31795.2635(20)[a] | 0.0004 | 31904.167(20)[b] | 0.0052 | 31935.825(20)[b] | -0.0026 |
| 5.5 | 31622.814(50)[c] | -0.0035 | 31597.025(20)[b] | -0.0083 | 31792.9301(20)[a] | 0.003 | 31795.8664(20)[a] | 0.0011 | 31933.062(50)[c] | 0.0437 | 31963.724(20)[b] | -0.0036 |
| 6.5 | 31595.716(20)[b] | -0.0081 | 31569.869(20)[b] | 0.0045 | 31793.9981(20)[a] | 0.0022 | 31796.6326(40)[a] | 0.0014 | 31961.611(20)[b] | -0.0035 | 31991.604(50)[c] | 0.0357 |
| 7.5 | 31568.883(50)[c] | 0.016 | 31543.004(20)[b] | 0.004 | 31795.1117(20)[a] | -0.0016 | 31797.5096(20)[a] | 0.0013 | 31989.954(50)[c] | -0.0008 | 32019.263(50)[c] | 0.024 |
| 8.5 | 31542.261(50)[c] | -0.0011 | 31516.438(20)[b] | 0.0016 | 31796.287(50)[c] | 0.0186 | 31798.4528(20)[a] | 0.0006 | 32018.009(50)[c] | 0.0011 | 32046.680(50)[c] | 0.0286 |
| 9.5 | 31515.922(50)[c] | -0.0009 | 31490.174(20)[b] | 0.0055 | 31797.4359(20)[a] | -0.0052 | 31799.4528(20)[b] | 0.033 | 32045.750(50)[c] | 0.02 | 32073.741(50)[c] | 0.0127 |
| 10.5 | 31489.870(50)[c] | 0.017 | 31464.198(20)[b] | 0.0106 | 31798.580(50)[c] | -0.0217 | 31800.376(50)[c] | 0.0098 | 32073.067(20)[b] | 0.001 | 32100.420(50)[c] | 0.0282 |
| 11.5 | 31464.070(50)[c] | 0.022 | 31438.485(50)[c] | 0.0058 | 31799.699(50)[c] | -0.015 | 31801.2443(40)[a] | 0.0019 | 32099.955(50)[c] | 0.0024 | 32126.583(50)[c] | 0.0184 |
| 12.5 | 31438.493(50)[c] | -0.0073 | 31413.12(10)[c] | 0.098 | 31800.717(50)[c] | -0.0195 | 31801.9945(40)[a] | -0.0014 | 32126.313(20)[b] | -0.0084 | 32152.13(10)[c] | -0.0359 |
| 13.5 | 31413.12(10)[c] | -0.068 | 31387.80(2)[c] | 0.0017 | 31801.65(10)[c] | 0.0327 | 31802.5662(40)[a] | -0.0013 | 32152.131(50)[c] | 0.0325 | 32177.18(10)[c] | 0.0709 |
| 14.5 | 31388.08(50)[c] | -0.023 | 31362.788(50)[c] | 0.023 | 31802.299(50)[c] | -0.0076 | 31802.8915(40)[a] | -0.0006 | 32177.183(50)[c] | -0.0185 | | |
| 15.5 | 31363.199(50)[c] | 0.0089 | 31337.924(50)[c] | 0.038 | | | 31802.8976(40)[a] | 0.0007 | | | | |
| 16.5 | 31338.425(50)[c] | -0.0006 | 31313.126(50)[c] | 0.018 | | | 31802.5022(40)[a] | 0.0013 | | | 32247.15(20)[d] | 0.0667 |
| 17.5 | 31313.772(50)[c] | 0.017 | 31288.402(50)[c] | 0.034 | | | 31801.6124(40)[a] | -0.0007 | 32247.47(20)[d] | -0.1003 | 32268.508(50)[c] | 0.0694 |
| 18.5 | 31289.117(50)[c] | 0.0018 | 31263.626(50)[c] | 0.032 | | | 31800.127(50)[c] | -0.0042 | 32269.089(50)[c] | 0.0508 | 32288.60(20)[d] | -0.0138 |
| 19.5 | 31264.449(50)[c] | 0.016 | 31238.742(50)[c] | 0.045 | | | 31797.965(50)[c] | 0.0249 | 32289.14(20)[d] | -0.1781 | 32307.55(20)[d] | 0.0752 |
| 20.5 | 31239.642(50)[c] | 0.020 | 31213.560(50)[c] | -0.010 | 31798.20(50)[c] | 0.0102 | | | 32308.240(50)[d] | -0.0361 | 32324.78(20)[d] | -0.0918 |
| 21.5 | 31214.683(50)[c] | 0.107 | 31188.04(20)[c] | -0.048 | | | | | 32325.75(20)[d] | -0.0133 | 32340.55(20)[d] | -0.087 |
| 22.5 | 31189.12(20)[d] | -0.049 | 31162.11(20)[c] | 0.010 | 31791.196(50)[c] | -0.0602 | 31785.678(50)[c] | -0.0167 | 32341.55(20)[d] | -0.0619 | 32354.53(20)[d] | -0.0499 |
| 23.5 | 31163.39(20)[d] | 0.14 | 31135.45(20)[c] | 0.025 | 31786.053(50)[c] | -0.0666 | 31779.069(50)[c] | -0.0672 | 32355.73(20)[d] | 0.0979 | 32366.42(20)[d] | -0.0625 |
| 24.5 | 31136.661(50)[c] | 0.023 | 31107.892(50)[c] | 0.043 | 31779.578(50)[c] | -0.0315 | 31770.930(50)[c] | -0.0421 | 32367.58(20)[d] | -0.0259 | 32376.03(20)[d] | -0.0622 |
| 25.5 | 31109.194(50)[c] | 0.075 | 31079.03(20)[d] | -0.079 | 31771.455(50)[c] | -0.0332 | 31760.845(50)[c] | -0.0757 | 32377.11(20)[d] | -0.1709 | 32383.11(20)[d] | -0.0041 |
| 26.5 | 31080.42(20)[d] | -0.011 | 31049.00(20)[d] | 0.11 | 31761.437(50)[c] | -0.039 | 31748.632(50)[c] | -0.017 | 32384.23(20)[d] | -0.1322 | 32387.24(20)[d] | 0.0392 |
| 27.5 | 31050.36(20)[d] | 0.10 | | | 31749.227(50)[c] | -0.0126 | 31733.814(50)[c] | 0.0571 | 32388.47(20)[d] | -0.0326 | | |
| 28.5 | | | | | 31734.444(50)[c] | 0.0648 | 31715.797(50)[c] | 0.0291 | | | | |
| 29.5 | | | | | 31716.486(50)[c] | 0.0699 | 31694.022(50)[c] | -0.0833 | | | | |
| 30.5 | | | | | 31694.771(50)[c] | -0.0048 | | | | | | |

a: Ubachs et al.[23], b: Bembenek et al. [24], c: This work, d: Heimer [21]

Satellite lines included in the fit: R$_{12ee}$(0.5): v= 31805.9006(20), obs-calc=0.003, ref: Ubachs et al.[23]. Q$_{12ef}$(0.5): v=31777.2704(60), obs-calc=0.0089, ref: Ubachs et al. [23].
P$_{12ee}$(2.5): v=31652.702(20), obs-calc=-0.0015, ref: Bembenek et al. [24].

Table 4: Measured transition frequencies of the 1-1 vibronic band of the $C^2\Sigma^+$-$X^2\Pi$ transition of CH (in cm$^{-1}$).

| J' | P$_{11ee}$ | obs-calc | P$_{22ff}$ | obs-calc | Q$_{11ef}$ | obs-calc | Q$_{22fe}$ | obs-calc | R$_{11ee}$ | obs-calc | R$_{22ff}$ | obs-calc |
|---|---|---|---|---|---|---|---|---|---|---|---|---|
| 0.5 | 31639.266(50)[a] | 0.011 | 31618.804(50)[a] | 0.007 | | | | | | | | |
| 1.5 | 31612.84(20)[b] | -0.116 | 31589.564(50)[a] | 0.045 | | | 31673.021(50)[a] | 0.007 | | | | |
| 2.5 | 31585.695(50)[a] | 0.015 | 31560.743(50)[a] | 0.044 | 31667.141(50)[a] | 0.010 | 31670.865(50)[a] | -0.009 | 31720.365(50)[a] | 0.050 | 31753.664(50)[a] | -0.002 |
| 3.5 | 31557.856(50)[a] | -0.047 | 31531.926(50)[a] | 0.026 | 31666.96(20)[a] | -0.001 | 31669.144(50)[a] | 0.003 | 31747.921(50)[a] | 0.025 | 31778.033(50)[a] | -0.029 |
| 4.5 | 31529.782(50)[a] | 0.013 | 31502.920(50)[a] | -0.076 | 31666.251(50)[a] | 0.014 | 31667.303(50)[a] | -0.033 | 31774.323(50)[a] | 0.028 | 31802.57(20)[b] | -0.061 |
| 5.5 | 31501.21(20)[b] | -0.133 | 31473.930(50)[a] | -0.008 | 31665.065(50)[a] | -0.002 | 31665.275(50)[a] | -0.016 | 31799.904(50)[a] | -0.002 | 31826.838(50)[a] | -0.015 |
| 6.5 | 31472.665(50)[a] | 0.015 | 31444.677(50)[a] | -0.013 | 31663.482(50)[a] | 0.014 | 31662.913(50)[a] | -0.001 | 31824.767(50)[a] | -0.027 | 31850.502(50)[a] | -0.020 |
| 7.5 | 31443.716(50)[a] | 0.018 | 31415.229(50)[a] | 0.004 | 31661.412(50)[a] | -0.014 | 31660.105(50)[a] | -0.021 | 31848.918(50)[a] | -0.021 | 31873.4812(50)[a] | -0.023 |
| 8.5 | 31414.478(50)[a] | -0.003 | 31385.47(50)[a] | -0.047 | 31658.884(50)[a] | -0.020 | 31656.74(20)[b] | -0.118 | 31872.310(50)[a] | 0.022 | 31895.61(20)[b] | -0.078 |
| 9.5 | 31384.99(50)[a] | 0.009 | 31355.509(50)[a] | -0.003 | 31655.857(50)[a] | 0.003 | 31653.018(50)[a] | -0.015 | 31894.770(50)[a] | 0.003 | 31916.992(50)[a] | 0.029 |
| 10.5 | 31355.17(50)[a] | -0.005 | 31325.191(50)[a] | 0.014 | 31652.08(20)[b] | -0.134 | 31648.571(50)[a] | 0.002 | 31916.301(50)[a] | 0.011 | 31937.185(50)[a] | -0.028 |
| 11.5 | 31325.052(50)[a] | 0.047 | 31294.480(50)[a] | 0.031 | 31647.913(50)[a] | 0.004 | 31643.363(50)[a] | -0.010 | 31936.62(20)[b] | -0.134 | 31956.341(50)[a] | 0.023 |
| 12.5 | 31294.478(50)[a] | 0.047 | 31263.13(20)[b] | -0.124 | 31642.890(50)[a] | 0.037 | 31637.339(50)[a] | 0.002 | 31956.074(50)[a] | 0.026 | 31974.113(50)[a] | -0.031 |
| 13.5 | 31263.13(20)[b] | -0.248 | 31231.528(50)[a] | 0.025 | 31636.952(50)[a] | 0.008 | 31630.340(50)[a] | 0.000 | 31974.114(50)[a] | 0.071 | 31990.570(50)[a] | 0.027 |
| 14.5 | 31231.708(50)[a] | -0.050 | 31199.10(50)[a] | 0.014 | 31630.086(50)[a] | 0.025 | 31622.20(20)[b] | -0.040 | 31990.570(50)[a] | -0.026 | 32005.371(50)[a] | 0.023 |
| 15.5 | 31199.465(50)[a] | 0.003 | 31165.74(20)[b] | -0.124 | 31622.05(20)[b] | -0.016 | 31612.79(20)[b] | -0.081 | 32005.532(50)[a] | -0.011 | 32018.22(20)[b] | -0.151 |
| 16.5 | 31166.18(20)[b] | -0.178 | 31131.56(20)[b] | -0.116 | 31612.64(20)[b] | -0.154 | 31602.07(20)[b] | 0.032 | 32018.58(20)[b] | -0.117 | 32029.14(20)[b] | -0.254 |
| 17.5 | 31132.11(20)[b] | -0.169 | 31096.08(20)[b] | -0.239 | 31601.97(20)[b] | -0.080 | 31589.54(20)[b] | 0.030 | 32029.50(20)[b] | -0.343 | 32037.96(20)[b] | -0.206 |
| 18.5 | 31096.85(20)[b] | -0.175 | 31059.37(20)[b] | -0.179 | 31589.564(50)[a] | -0.044 | 31575.17(10)[a] | 0.157 | 32038.52(20)[b] | -0.210 | 32044.37(20)[b] | -0.025 |
| 19.5 | 31060.26(20)[b] | -0.093 | 31021.05(20)[b] | -0.019 | 31575.208(50)[a] | 0.016 | 31558.290(50)[a] | 0.057 | 32045.04(20)[b] | -0.029 | 32047.74(20)[b] | 0.000 |
| 20.5 | 31021.86(20)[b] | -0.107 | 30980.42(20)[b] | -0.103 | 31558.528(50)[a] | 0.043 | | | 32048.51(20)[b] | -0.007 | 32047.83(20)[b] | 0.031 |
| 21.5 | 30981.38(20)[b] | -0.130 | | | | | | | 32048.74(20)[b] | 0.065 | 32044.22(20)[b] | 0.119 |
| 22.5 | | | | | | | | | 32045.19(20)[b] | 0.121 | 32036.01(20)[b] | -0.077 |
| 23.5 | | | | | | | | | 32036.88(20)[b] | -0.265 | | |

a: from this work  b: From Li et al. [25]

Table 5: Measured transition frequencies of the 2-2 vibronic band of the $C^2\Sigma^+$-$X^2\Pi$ transition of CH (in cm$^{-1}$).

| J' | P$_{11ee}$ | obs-calc | P$_{22ff}$ | obs-calc | Q$_{11ef}$ | obs-calc | Q$_{22fe}$ | obs-calc | R$_{11ee}$ | obs-calc | R$_{22ff}$ | obs-calc |
|---|---|---|---|---|---|---|---|---|---|---|---|---|
| 0.5 | 31403.265(50) | | 31382.697(50) | -0.046 | 31421.13(20)[a] | 0.21 | | | | | | |
| 1.5 | 31377.127(50) | -0.039 | 31353.074(50) | 0.034 | 31428.466(50) | -0.009 | 31433.337(50) | 0.029 | 31446.194(50) | -0.048 | | |
| 2.5 | 31349.393(50) | 0.016 | 31322.995(50) | 0.016 | 31427.713(50) | 0.023 | 31428.887(50) | -0.019 | 31478.838(50) | 0.001 | 31508.517(50) | 0.023 |
| 3.5 | 31320.331(50) | 0.017 | 31292.060(50) | -0.062 | 31425.182(50) | 0.018 | 31424.083(50) | 0.024 | 31502.91(20)[a] | -0.077 | 31528.732(50) | -0.045 |
| 4.5 | 31290.111(50) | 0.017 | 31260.337(50) | 0.019 | | | 31418.327(50) | 0.028 | 31525.149(50) | -0.036 | 31548.341(50) | -0.028 |
| 5.5 | 31258.774(50) | 0.026 | 31227.402(50) | -0.076 | 31416.163(50) | 0.024 | 31411.463(50) | 0.045 | 31545.792(50) | 0.019 | 31566.746(50) | 0.017 |
| 6.5 | 31226.285(50) | 0.026 | | | 31409.706(50) | 0.022 | 31403.265(50) | -0.001 | 31564.770(50) | -0.002 | 31583.575(50) | -0.030 |
| 7.5 | | | | | 31401.861(50) | 0.007 | 31393.696(50) | -0.004 | 31582.099(50) | -0.006 | 31598.779(50) | -0.014 |
| 8.5 | | | 31121.790(50) | 0.003 | 31392.48(20) | -0.062 | 31382.528(50) | -0.025 | 31597.662(50) | 0.017 | | |
| 9.5 | 31121.298(50) | 0.0003 | 31083.782(50) | 0.036 | 31381.636(50) | 0.030 | 31369.645(50) | 0.009 | 31611.1787(50) | -0.046 | | |
| 10.5 | 31083.432(50) | -0.007 | | | 31368.853(50) | -0.013 | 31354.743(50) | 0.022 | | | 31632.014(50) | -0.056 |
| 11.5 | | | 31002.324(50) | -0.019 | 31354.144(50) | 0.039 | 31337.539(50) | -0.0004 | 31631.650(50) | -0.001 | | |
| 12.5 | 31002.324(50) | -0.024 | | | 31337.040(50) | -0.019 | | | | | | |
| 13.5 | | | | | | | | | 31641.298(50) | 0.057 | | |
| 14.5 | | | | | | | | | | | | |
| 15.5 | | | | | | | | | | | 31628.288(50) | 0.058 |
| 16.5 | | | | | | | | | 31628.507(50) | -0.064 | | |

a: From Herzberg and Johns [22]; all other lines from this work